\documentclass[reprint,unsortedaddress,amsmath,amssymb,aps,floatfix]{revtex4-2}

\usepackage{graphicx}
\usepackage{dcolumn}
\usepackage{bm}
\usepackage{amsmath,amssymb,bm}
\usepackage{mathtools}
\usepackage{color}
\usepackage{bbold}

\newcommand{\Tr}{\operatorname{Tr}}
\newcommand{\rd}{\mathrm{d}}

\newcommand{\Mmat}{\mathbb{M}}
\newcommand{\Sr}{\Sigma_{r}}
\newcommand{\Ss}{\Sigma_{s}}

\begin{document}
\title{Dynamics and Statistical Mechanics of Born–Oppenheimer Auxiliary Velocities}

\author{Paula Sierra}

\affiliation{
 Centre Européen de Calcul Atomique et Moléculaire (CECAM), Ecole Polytechnique Fédérale de Lausanne, 1015 Lausanne, Switzerland\\
}

\author{Benjamin Rotenberg}
\affiliation{
 Physicochimie des \'Electrolytes et Nanosyst\`emes Interfaciaux, Sorbonne Universit\'e, CNRS, Paris, 75005, France
}
\affiliation{Réseau sur le Stockage Electrochimique de l’Energie (RS2E), FR CNRS 3459, 80039 Amiens Cedex, France}
\author{Sara Bonella}
\email{sara.bonella@epfl.ch}
\affiliation{
 Centre Européen de Calcul Atomique et Moléculaire (CECAM), Ecole Polytechnique Fédérale de Lausanne, 1015 Lausanne, Switzerland\\
}

\begin{abstract}
In Born–Oppenheimer (BO) dynamics, auxiliary variables such as electron densities, induced polarization and electrode charges follow stationarity conditions, yet their time derivatives, such as currents, are physical observables. Using Mass-Zero constrained dynamics, we interpret these derivatives as explicit dynamical variables and derive the invariant measure giving their equilibrium statistical properties. We illustrate the framework in constant-potential molecular dynamics, where the electrode-charge velocities reproduce BO current fluctuations and the frequency-dependent admittance.
\end{abstract}

\maketitle

{\bf Introduction} In molecular dynamics, many physically relevant quantities
are fixed by instantaneous stationarity conditions rather than by their own
equations of motion. Examples include electron densities in orbital-free~\cite{pearson:1993} and Kohn-Sham
electronic-structure dynamics~\cite{marx:2012-book}, induced polarization in classical force fields~\cite{dick:1958,jacucci:1976,wilson:1993,wilson:1996b},
and electrode charges in constant-potential simulations~\cite{Siepmann1995}.
In the Born--Oppenheimer (BO) description these auxiliary variables are
functions of the evolving physical coordinates. In particular, their
time-derivatives (referred to as ``auxiliary velocities'' hereafter) follow from the chain rule, but as derived quantities, without a
statistical-mechanical framework of their own. These derivatives, however, correspond to physical observables. For example, polarization and electrode currents determine transport and, through fluctuation--dissipation relations, frequency-dependent response~\cite{Pireddu2023,Pireddu2024,Kivelson1975,Caillol1989a,Caillol1989b,Sega2013}.
A consistent dynamical and statistical-mechanical description of these derivatives is therefore interesting and useful.

A dynamical framework is introduced in extended-Lagrangian methods that propagate auxiliary variables with fictitious
inertia, as in Car--Parrinello dynamics~\cite{car:1985,marx:2012-book}, or combine auxiliary
dynamics with approximate self-consistent updates~\cite{kolafa:2004,kuhne:2007,niklasson:2006,niklasson:2021}. Finite values of the fictitious electronic mass, however, are known to introduce systematic deviations from Born–Oppenheimer dynamics~\cite{Tangney:2002,10.1063/1.2162893}, including effective renormalization of the ionic masses and associated biases in dynamical observables such as vibrational frequencies~\cite{10.1021/ct6001913,10.1039/c0cp00411a}.
Mass-Zero (MaZe) constrained dynamics~\cite{Bonella2020,Coretti2020} removes these spurious effects via an extended dynamical system that
enforces a rigorous null inertia limit and imposes the stationarity conditions as holonomic constraints throughout the
propagation. MaZe has been applied to polarizable
models~\cite{Coretti2018}, orbital-free density functional
theory~\cite{Coretti2022ofdft}, shell models~\cite{Girardier2021} and
constant-potential electrodes~\cite{Coretti2020}, and its mass-zero limit is
proven to reproduce BO dynamics and equilibrium statistics for the physical
variables. Here we focus on the auxiliary variables.

We show that (i) in Hamiltonian variables the auxiliary momenta and the
constraint multipliers scale linearly with the inertia, so that the mass-zero
limit is regular and the auxiliary velocities remain finite and explicit;
(ii) these velocities coincide exactly with the time derivatives of the BO
stationary solutions along the same physical trajectory; and (iii) the
phase-space compressibility and invariant measure of the constrained flow
provide the statistical description needed for equilibrium averages and
time-correlation functions of the velocities. We illustrate the framework on
constant-potential electrodes, where the auxiliary velocities are electrical
currents, by comparing MaZe with a BO reference for the current,
its autocorrelation, and the frequency-dependent admittance.

\vspace{1ex}
{\bf Theory} We consider a system of $N$ physical particles with
coordinates $r$, conjugate momenta $p$, and masses
collected in the diagonal matrix $\Mmat$, coupled to $f$
auxiliary variables $s$ through a (possibly generalised) potential 
$V(r,s)$. In the Born--Oppenheimer (BO) description,
the auxiliary variables are determined at each physical
configuration by the stationarity conditions
\begin{equation}
\sigma_\alpha(r,s)
\equiv \frac{\partial V(r,s)}{\partial s_\alpha}=0,
\qquad \alpha=1,\ldots,f.
\label{eq:stationarity}
\end{equation}
We assume that these conditions define a smooth,
locally unique solution $s^*(r)$ so that the
corresponding $(f\times f)$ Jacobian  with respect to the auxiliary
variables, $(\Sigma_s)_{\alpha\beta}
=\frac{\partial\sigma_\alpha}{\partial s_\beta}$, is nonsingular.

MaZe dynamics enforces
Eq.~\eqref{eq:stationarity} as a set of holonomic
constraints while propagating the auxiliary coordinates.
To formulate the dynamics in Hamiltonian variables,
we initially assign a finite inertia $\mu$ to the
auxiliary variables and introduce their conjugate
momenta $\pi$. The extended Hamiltonian is
\begin{equation}
H(r,p,s,\pi)
=\frac{1}{2}p^{\mathsf T}\Mmat^{-1}p
+\frac{1}{2\mu}\pi^{\mathsf T}\pi
+V(r,s).
\label{eq:H}
\end{equation}
The constrained equations of motion are
\begin{equation}
\label{eq:EOMfinite}
\begin{aligned}
\dot r &= \Mmat^{-1}p,\\
\dot p &= -\nabla_r V-\Sigma_r^{\mathsf T}\lambda,\\
\dot s &= \pi/\mu,\\
\dot\pi &= -\Sigma_s^{\mathsf T}\lambda,
\end{aligned}
\end{equation}
where the $f$-dimensional $\lambda$ denotes the Lagrange multipliers and we have introduced the $(f\times 3N)$ matrix $(\Sigma_r)_{\alpha i}
=\frac{\partial\sigma_\alpha}{\partial r_i}$.  The auxiliary potential forces vanish on the
constraint manifold because
$\nabla_s V=\sigma=0$ (see Eq.~\eqref{eq:stationarity}).

Preservation of the stationarity conditions requires
the velocity constraint
\begin{equation}
\dot\sigma
=\Sigma_r\Mmat^{-1}p+\mu^{-1}\Sigma_s\pi=0.
\label{eq:sigmadot}
\end{equation}
This relation gives
\begin{equation}
\pi=-\mu\Sigma_s^{-1}\Sigma_r\Mmat^{-1}p
\equiv\mu\hat\pi,
\qquad
\dot s=\hat\pi.
\label{eq:scaledmomentum}
\end{equation}
Thus, the auxiliary momenta scale linearly with
$\mu$, whereas the auxiliary velocities remain finite for all values of the fictitious inertia.

Differentiating the constraints once more and
using Eqs.~\eqref{eq:EOMfinite}, we obtain a linear equation for the
multipliers, whose solution can be written as
\begin{equation}
\lambda=\mu\gamma,
\qquad
\gamma=\tilde{\mathbb{Z}}^{-1}F,
\qquad
\tilde{\mathbb{Z}}=
\Sigma_s\Sigma_s^{\mathsf T}
+\mu\Sigma_r\Mmat^{-1}\Sigma_r^{\mathsf T},
\label{eq:lambda}
\end{equation}
where
\begin{equation}
F_\alpha=
\sum_{a,b}
\frac{\partial^2\sigma_\alpha}
{\partial x_a\partial x_b}
\dot x_a\dot x_b
-\left(\Sigma_r\Mmat^{-1}\nabla_rV\right)_\alpha,
\label{eq:Fdef}
\end{equation}
with $x=(r,s)$ and
$\dot x=(\Mmat^{-1}p,\hat\pi)$, which is finite for $\mu\to0$ by
Eq.~\eqref{eq:scaledmomentum} ($\hat\pi=\pi/\mu$).
For nonsingular $\Sigma_s$ and finite constrained
velocities, $\tilde{\mathbb{Z}}$ remains invertible and
$\gamma$ remains finite as $\mu\to0$.
Consequently, the constraint multipliers also
scale linearly with the auxiliary inertia.

Taking the mass-zero limit while retaining the
scaled momenta $\hat\pi$ gives
\begin{equation}
\label{eq:EOMzero}
\begin{aligned}
\dot r &= \Mmat^{-1}p,\\
\dot p &= -\nabla_rV(r,s),\\
\dot s &= \hat\pi,\\
\dot{\hat\pi} &= -\Sigma_s^{\mathsf T}\gamma_0,
\end{aligned}
\end{equation}
subject to the position and velocity constraints and where $\gamma_0$ is the zero inertia Lagrange multiplier.
Since $s=s^*(r)$, the equations for the physical
variables reduce to
\begin{equation}
\dot r=\Mmat^{-1}p,
\qquad
\dot p=-\nabla_rV(r,s^*(r)).
\label{eq:BOdynamics}
\end{equation}
These are the BO equations of motion. For identical
initial physical conditions, MaZe and BO therefore
generate the same physical trajectories in the
mass-zero limit.

We now establish the correspondence for the
auxiliary velocities. In BO dynamics, the stationary
solution $s^*(r)$ follows the physical trajectory
without possessing an independent equation of motion.
Its time derivative is determined by the chain rule,
\begin{equation}
\frac{d}{dt}s^*(r(t))
=\frac{\partial s^*}{\partial r}\dot r.
\label{eq:BOchain}
\end{equation}
Differentiating the identity
$\sigma(r,s^*(r))=0$ with respect to $r$ yields
$\Sigma_r+\Sigma_s\,\partial s^*/\partial r=0$,
and therefore
\begin{equation}
\frac{d}{dt}s^*(r(t))
=-\Sigma_s^{-1}\Sigma_r\Mmat^{-1}p.
\label{eq:BOvelocity}
\end{equation}
Comparison with Eq.~\eqref{eq:scaledmomentum}
shows that the time derivative of the BO
stationary solution coincides with the constrained
auxiliary velocity,
\begin{equation}
\frac{d}{dt}s^*(r(t))=\hat\pi(t)=\dot s(t),
\label{eq:velocitycorrespondence}
\end{equation}
when evaluated along the same physical trajectory.
This identity follows from the geometry
of the stationary manifold and holds for any
finite $\mu$. At finite inertia, the constrained
and BO physical trajectories generally differ
because of the constraint force acting on the
physical momenta. In the mass-zero limit, this
force vanishes and the physical trajectories
coincide, so the correspondence extends to the
time evolution as well.

\begin{figure}[t]
  \centering
  \includegraphics[width=0.5\textwidth]{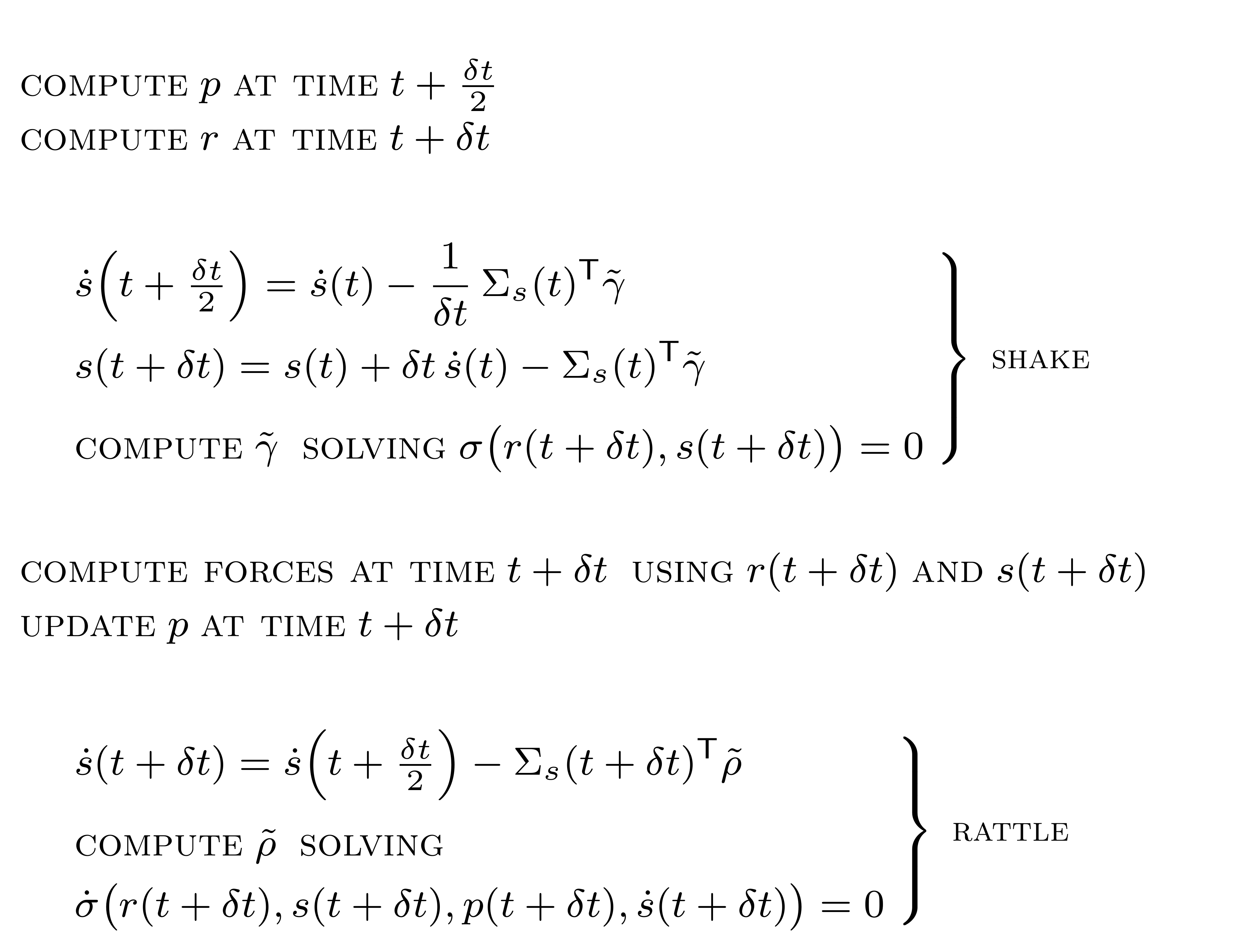}
  \caption{Integration cycle of the mass-zero constrained dynamics.
  Starting from $\{r,\dot r,s,\dot s\}(t)$, a velocity half-step is
  followed by the \textsc{shake} update enforcing
  $\sigma(t+\delta t)=0$; after recomputing the forces, the
  \textsc{rattle} projection enforces
  $\dot\sigma(t+\delta t)=0$ [Eq.~\eqref{eq:sigmadot}].}
  \label{fig:algorithm}
\end{figure}

The identity is a direct consequence of the geometry of the stationary manifold
(trivial, in that sense), yet it is useful: in MaZe,
$\hat\pi$ is a regular phase-space variable of a well-defined dynamical
system, produced directly by the constrained integrator detailed in Fig.~\ref{fig:algorithm}, without
differentiating the stationary solution~\footnote{We note that previous solutions of MaZe dynamics employed a Verlet algorithm for the auxiliary variables, thus requiring only SHAKE to solve the constraints. The focus on the auxiliary velocities requires the velocity Verlet, combined with SHAKE and RATTLE, integrator introduced in this work.}. The equilibrium fluctuations of the auxiliary coordinates and velocities, require a
consistent statistical description of the extended constrained
dynamics. This can be conveniently obtained within the formalism of non-Hamiltonian statistical mechanics~\cite{Tuckerman1999,Tuckerman2001}. Eliminating the
constraint multipliers generally leads to a
compressible flow in the ambient phase-space
variables $(r,p,s,\pi)$. As shown in the
Supplemental Material (SM)~\cite{SM}, its compressibility is
\begin{equation}
\kappa_\mu
=-\operatorname{Tr}
\left(\tilde{\mathbb{Z}}^{-1}\dot{\tilde{\mathbb{Z}}}\right)
=-\frac{d}{dt}\ln|\tilde{\mathbb{Z}}|.
\label{eq:compressibility}
\end{equation}
where $|\tilde{\mathbb{Z}}|$ denotes the determinant of $\tilde{\mathbb{Z}}$.
The corresponding invariant metric factor can
be chosen as
\begin{equation}
\sqrt g=|\tilde{\mathbb{Z}}|,
\label{eq:metric}
\end{equation}
up to a constant factor, which cancels in normalized averages. The constrained equilibrium average of an
observable $A$ is then
\begin{equation}
\begin{aligned}
\langle A\rangle_\mu=\frac{1}{\mathcal N_\mu}
\int dr\,dp\,ds\,d\pi\; \Big \{
|\tilde{\mathbb{Z}}|\,P_\mu(r,p,s,\pi)\,
\delta(\sigma)\,\\
\times \delta(\dot\sigma)\,A(r,p,s,\pi)\Big \},
\end{aligned}
\label{eq:average}
\end{equation}
where $\mathcal N_\mu=\int dr\,dp\,ds\,d\pi\,|\tilde{\mathbb{Z}}|\,P_\mu(r,p,s,\pi)\,\delta(\sigma)\,\delta(\dot\sigma)$
is the normalization factor, $P_\mu(r,p,s,\pi)$ is the stationary density of the extended dynamics, the delta functions enforce the constraints and the velocity constraints, and
$A(r,p,s,\pi)$ is the observable.

To obtain a regular statistical description in
the mass-zero limit, we introduce the scaled
auxiliary momenta $\pi=\mu\hat\pi$. In these variables, the velocity
constraint becomes
\begin{equation}
\delta(\dot\sigma)=\delta(\Sigma_s\hat\pi+\Sigma_r\Mmat^{-1}p)
=|\Sigma_s|^{-1}\,\delta\!\big(\hat\pi+\Sigma_s^{-1}\Sigma_r\Mmat^{-1}p\big),
\end{equation}
with no explicit dependence on $\mu$. The momentum Jacobian $\mu^{f}$ cancels
between numerator and denominator of the
normalized average, and the Hamiltonian becomes
$H_\mu=\frac12p^{\mathsf T}\Mmat^{-1}p+\frac{\mu}{2}\hat\pi^{\mathsf T}\hat\pi+V(r,s)$,
with the regular limit
$H_0=\frac12p^{\mathsf T}\Mmat^{-1}p+V(r,s)$.
Since $\tilde{\mathbb{Z}}\to\Sigma_s\Sigma_s^{\mathsf T}$, the limiting average is
\begin{equation}
\begin{aligned}
\langle A\rangle_0=\frac{1}{\mathcal N_0}
\int dr\,dp\,ds\,d\hat\pi\; \Big \{
|\Sigma_s\Sigma_s^{\mathsf T}|\,P_0(r,p,s,\hat\pi)\,
\delta(\sigma)\,\\
\times
\delta(\Sigma_s\hat\pi+\Sigma_r\Mmat^{-1}p)\,A(r,p,s,\hat\pi)\Big \},
\end{aligned}
\label{eq:averagezero}
\end{equation}
where $\mathcal N_0$ is the corresponding normalization and $P_0(r,p,s,\hat\pi)$ is the limiting
stationary density.
Marginalization over $s$ and $\hat{\pi}$, shows that
the physical variables are distributed according to $P_0(r,p,s^*(r))$,
the BO density for the chosen ensemble. This
recovers the known BO limit of the constrained measure~\cite{Bonella2020,Coretti2020}. The derivation of this limit in the present variables is given in the SM~\cite{SM}.

In the full extended system, the velocity $\hat\pi$ is slaved to $p$, but remains an explicit phase-space variable.
Together with the dynamical correspondence above,
the measure allows equilibrium time-correlation functions
and linear-response relations to be formulated
directly for observables involving the auxiliary velocities~\cite{10.1063/1.3354126}.

\begin{figure}[!t]
  \centering
  \includegraphics[width=\linewidth]{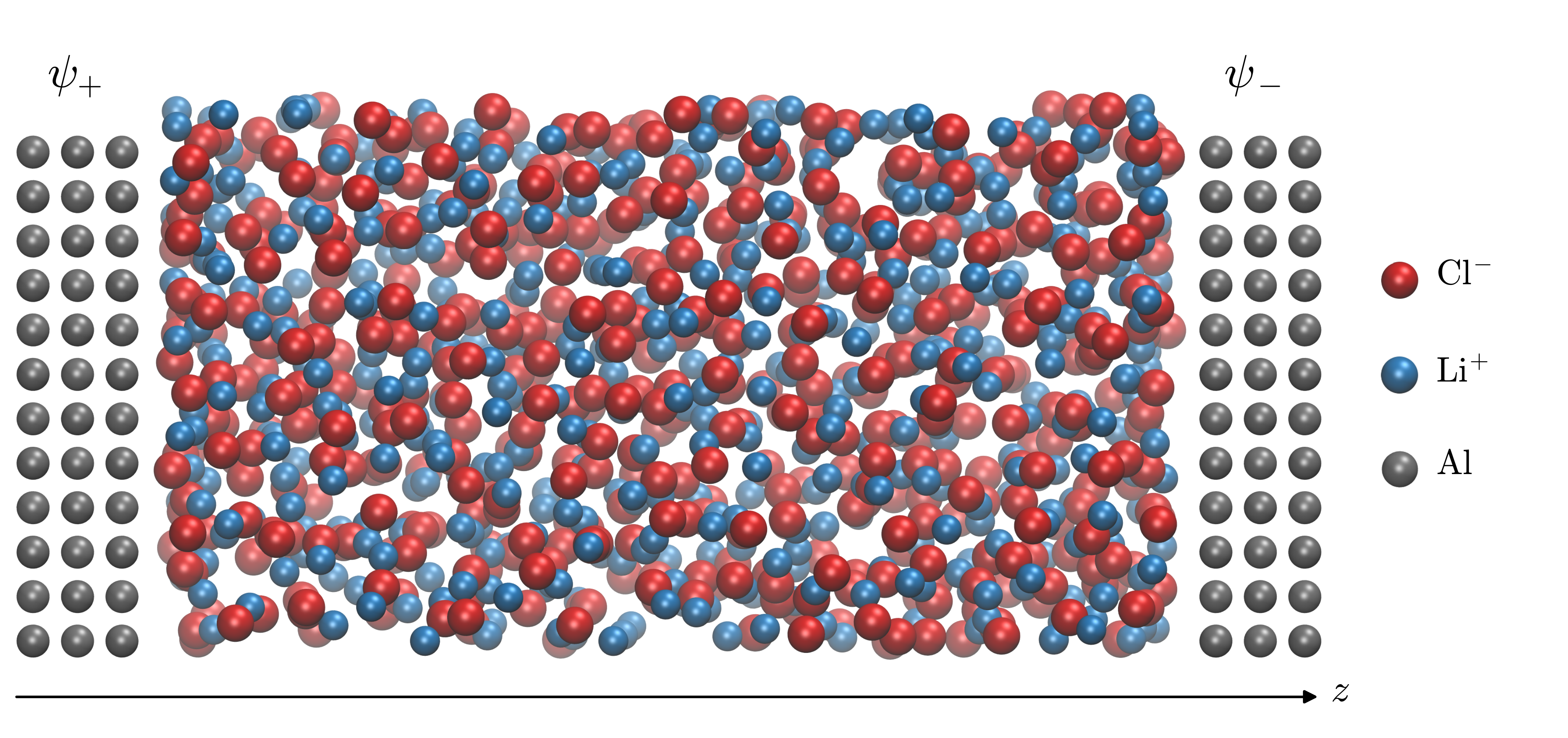}
  \caption{Snapshot of molten LiCl (Li$^+$ blue, Cl$^-$ red) between two Al electrodes (grey) held at constant potentials $\psi_+$ and $\psi_-$, with periodic boundary conditions in $x$ and $y$. The electrode charges fluctuate under the constant-potential and electroneutrality conditions.
  }
  \label{fig:schemeLiCl}
\end{figure}

\vspace{1ex}
{\bf Application to constant-potential electrodes}
We illustrate the correspondence using the fluctuating-charge
model of constant-potential electrodes, represented in Fig.~\ref{fig:schemeLiCl}. The system is an electrolyte of $N$ ions, whose positions $r=\{{r}_i\}^{3N}_{i=1}$ are the physical degrees of freedom, confined between two metallic electrodes, labelled $+$ and $-$, each composed of fixed atoms. Every
electrode atom $\alpha$ carries a Gaussian charge distribution of fixed width and fluctuating magnitude $Q_\alpha$ (see SM~\cite{SM}), and the sites of each electrode are held at an imposed potential $\psi_\alpha$ ($\psi_+$ and $\psi_-$ for the two electrodes, as indicated in the scheme). The $M$ electrode-site charges, which adjust
to the ionic configuration so as to keep each site at its imposed potential, are the auxiliary coordinates, with an additional auxiliary variable $\nu$ needed to enforce global electroneutrality:
\begin{equation}
 s=(Q_1,\ldots,Q_M,\nu), \qquad f=M+1.
 \label{eq:electrode_variables}
\end{equation}
The physical coordinates $r$ evolve based on the negative gradient of
\begin{equation}
 V(r,s)=U(r)+U_C(r,Q)-\sum_{\alpha=1}^{M}\psi_\alpha Q_\alpha
          +\nu\sum_{\alpha=1}^{M}Q_\alpha,
 \label{eq:electrode_potential}
\end{equation}
where $U(r)$ denotes the non-Coulomb interactions of the electrolyte, $U_C$ is the electrostatic energy of the electrolyte
and electrode charges (specified in the SM~\cite{SM}). The third term in Eq.~\eqref{eq:electrode_potential} is the reversible work to charge the electrodes, while the fourth imposes the electroneutrality condition via the Lagrange multiplier $\nu$, and $\psi_\alpha$ is the imposed potential at electrode site
$\alpha$. The BO stationarity conditions read
\begin{equation}
 \frac{\partial V}{\partial Q_\alpha}=0\quad(\alpha=1,\ldots,M),
 \qquad \frac{\partial V}{\partial\nu}
 =\sum_{\alpha=1}^{M}Q_\alpha=0.
 \label{eq:electrode_constraints}
\end{equation}
They impose the constant-potential conditions and overall
electrode electroneutrality. In BO dynamics, the stationary
charges are obtained by minimization at each ionic configuration;
in MaZe, the same conditions are maintained as holonomic
constraints during propagation.

In this model, $\Sigma_s$ is the (bordered) Hessian of $V$ with respect to the
electrode charges and $\nu$, which depends only on the fixed electrode geometry and not on
the ionic coordinates. Consequently, the
mass-zero compressibility vanishes,
$\kappa_0=-\frac{d}{dt}\ln|\Sigma_s\Sigma_s^{\mathsf T}|=0$.
This property simplifies the statistical description but is
not required for the general BO--MaZe correspondence.

Let $Q_+$ and $Q_-$ denote the total charges of the two
electrodes. Since $Q_++Q_-=0$, the capacitor charge and
its associated current are
\begin{equation}
 Q(t)=\frac{Q_+(t)-Q_-(t)}{2},\qquad I(t)=\dot Q(t).
 \label{eq:capacitor_current}
\end{equation}
In MaZe, $I(t)$ is obtained directly from the constrained
charge velocities, which are components of $\hat\pi$.
In BO dynamics, it is the derivative of the BO stationary
charge $Q^*(r(t))$ along the ionic trajectory, evaluated in the following by central
finite differences. The velocity
correspondence identifies these currents.

The frequency-dependent admittance $Y(\omega)$, which is the linear response function relating the current $I$ flowing through the cell to a small oscillating voltage applied between the two electrodes~\cite{Pireddu2023,Pireddu2024}, can be computed from equilibrium fluctuations through two expressions~\cite{Pireddu2023}. The current-based Nyquist--Johnson
relation is
\begin{equation}
 Y(\omega)=\beta\int_0^\infty
 \langle\delta I(0)\,\delta I(t)\rangle
 e^{-i\omega t}\,dt,
 \label{eq:Ycurrent}
\end{equation}
while its charge-correlation counterpart is
\begin{equation}
 Y(\omega)=\beta\left[
 i\omega\langle\delta Q^2\rangle
 +\omega^2\int_0^\infty
 \langle\delta Q(0)\,\delta Q(t)\rangle
 e^{-i\omega t}\,dt\right].
 \label{eq:Ycharge}
\end{equation}
Here $\beta=(k_{\mathrm B}T)^{-1}$,
$\delta Q=Q-\langle Q\rangle$, and
$\delta I=I-\langle I\rangle$. Eq.~\eqref{eq:Ycurrent} probes the auxiliary velocities directly,
whereas Eq.~\eqref{eq:Ycharge} depends only on the
stationary auxiliary coordinates. Comparing both estimators
in MaZe and BO therefore tests the dynamical and statistical
correspondence at the level of a physical response function.
The corresponding impedance is $Z(\omega)=1/Y(\omega)$.

In our simulations, we consider molten LiCl between two aluminum electrodes simulated with MetalWalls \cite{MetalWalls2020, MetalWalls2022}. Both electrodes are held at the same null potential, $\psi_\alpha=0$~V. The cell contains 500 Li$^+$
and 500 Cl$^-$ ions and two electrodes of 216 fixed Al
atoms each, with $L_x=L_y=24.30\,\text{\AA}$ and $L_z=57.93\,\text{\AA}$. Periodic boundary conditions are applied parallel to the electrodes. The distance between the innermost electrode planes is $L=49.83$~\AA. The simulation protocol and interaction models are described in the
SM~\cite{SM}. We compare MaZe (integrated via the algorithm detailed in Fig.~\ref{fig:algorithm}) with a BO
reference in which the electrode charges are minimized
at every time step using conjugate-gradient.

\begin{figure}[!t]
  \centering
  \includegraphics[width=0.85\linewidth]{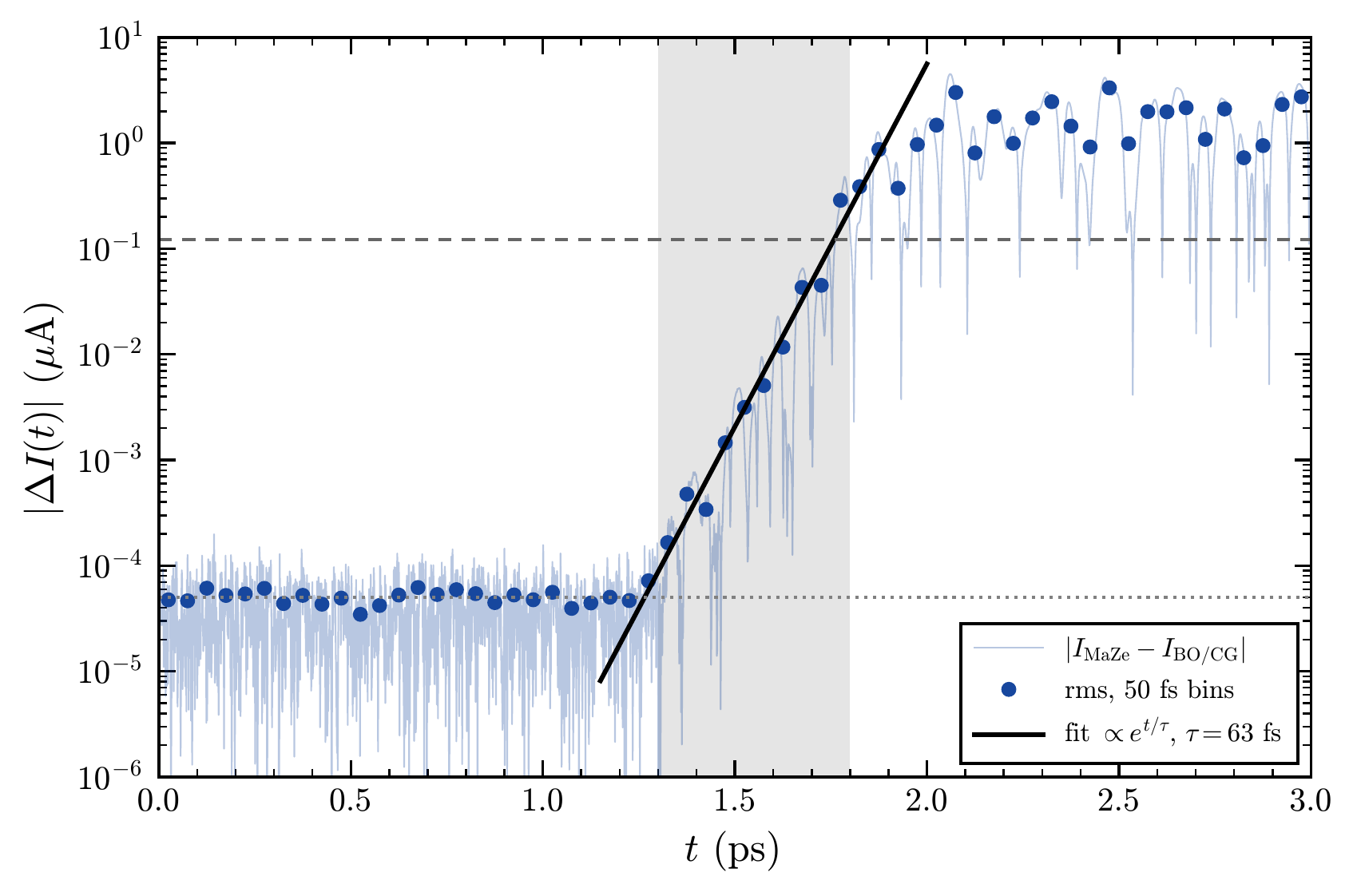}
  \caption{Difference between the electrode currents $I=\dot{Q}$ from MaZe and BO/CG for the same initial configuration: $|I_{\rm MaZe}-I_{\rm BO/CG}|$ (thin line) and its rms in 50~fs bins (dots) over the first 3~ps. The difference is flat ($5\times10^{-5}~\mu$A, dotted line) up to $\simeq1.2$~ps, then grows exponentially as $e^{t/\tau}$ with $\tau\approx63$~fs (shaded fit window) and saturates at $\simeq\sqrt{2}\,$rms$(I)$, i.e. the trajectories decorrelate. The dashed line marks $0.1\times$ rms$(I)$.
  }
  \label{fig:dsdt_maze_vs_bo}
\end{figure}

\begin{figure}[!t]
  \centering
    \includegraphics[width=0.85\linewidth]{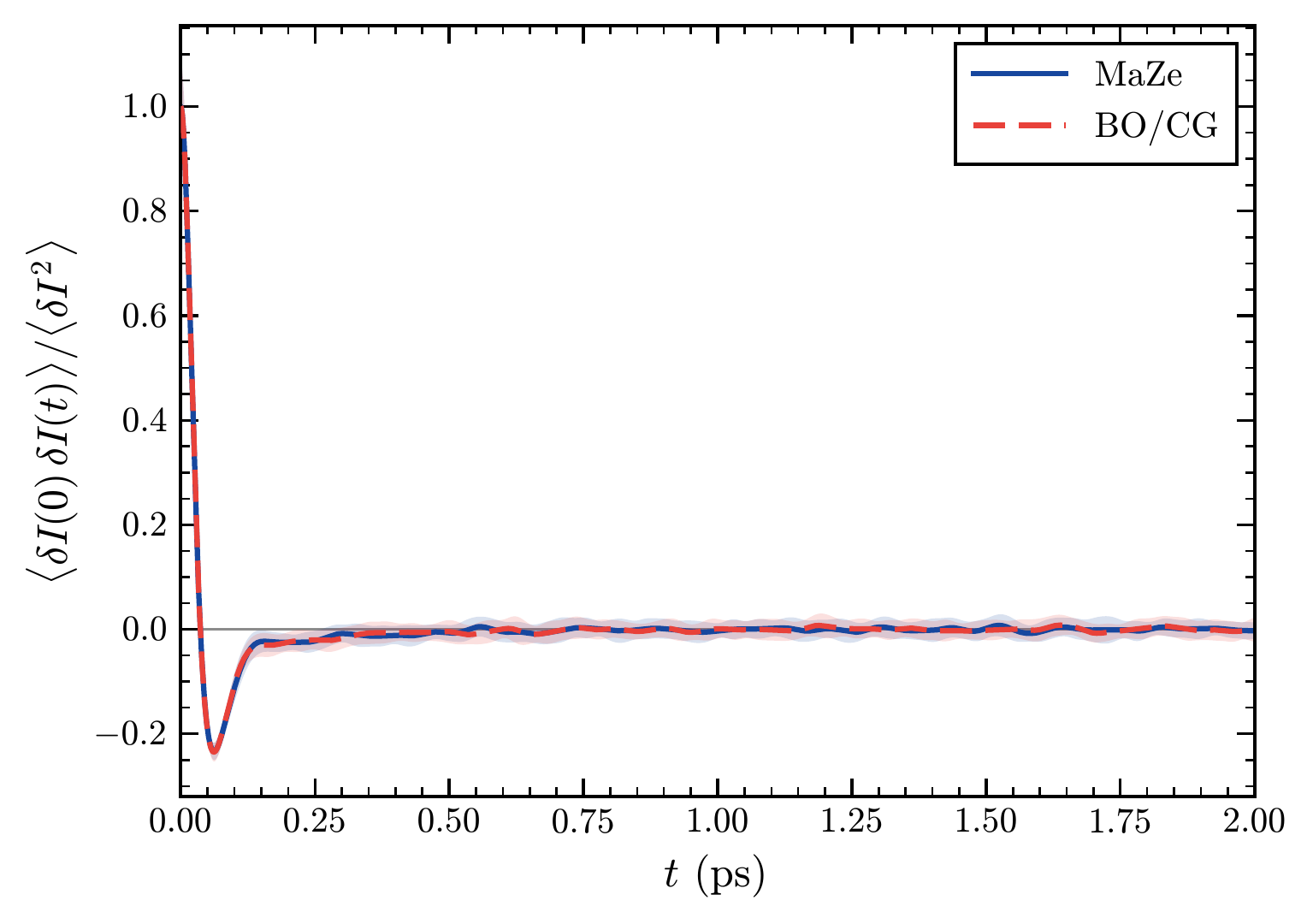}
  \caption{
Normalized autocorrelation function of the current fluctuations, $\langle\delta I(0)\,\delta I(t)\rangle$, obtained from MaZe dynamics (solid blue) and from the BO/CG reference (dashed red). The shaded areas indicate $\pm1$ standard deviation over non-overlapping 100~ps blocks.}
  \label{fig:acf}
\end{figure}

Fig.~\ref{fig:dsdt_maze_vs_bo} compares the electrode currents from MaZe charge velocities and from central finite differences of the BO stationary charges, starting from the same initial conditions. For the first $1.2$~ps the difference stays at a constant plateau of $5\times10^{-5}~\mu$A ($\sim4\times10^{-5}$ of the rms current). It then grows exponentially as $e^{t/\tau}$, with $\tau\approx63$~fs, reaches $10\%$ of the rms current at $1.76$~ps, and saturates at $\simeq\sqrt{2}$ times the rms current, the value for two uncorrelated signals. The separation time is nearly independent of the timestep ($1.76$~ps for $1$~fs and $1.81$~ps for $0.25$~fs), so it is not set by the integration error but by the onset of the (Lyapunov-like) amplification of small numerical differences by the chaotic dynamics of the liquid. Fig.~\ref{fig:acf} compares the equilibrium current autocorrelation functions: MaZe and BO agree within statistical uncertainty as the divergence of the currents affects times longer than the correlation decay.

\begin{figure*}[!t]
  \centering
    \includegraphics[width=0.85\linewidth]{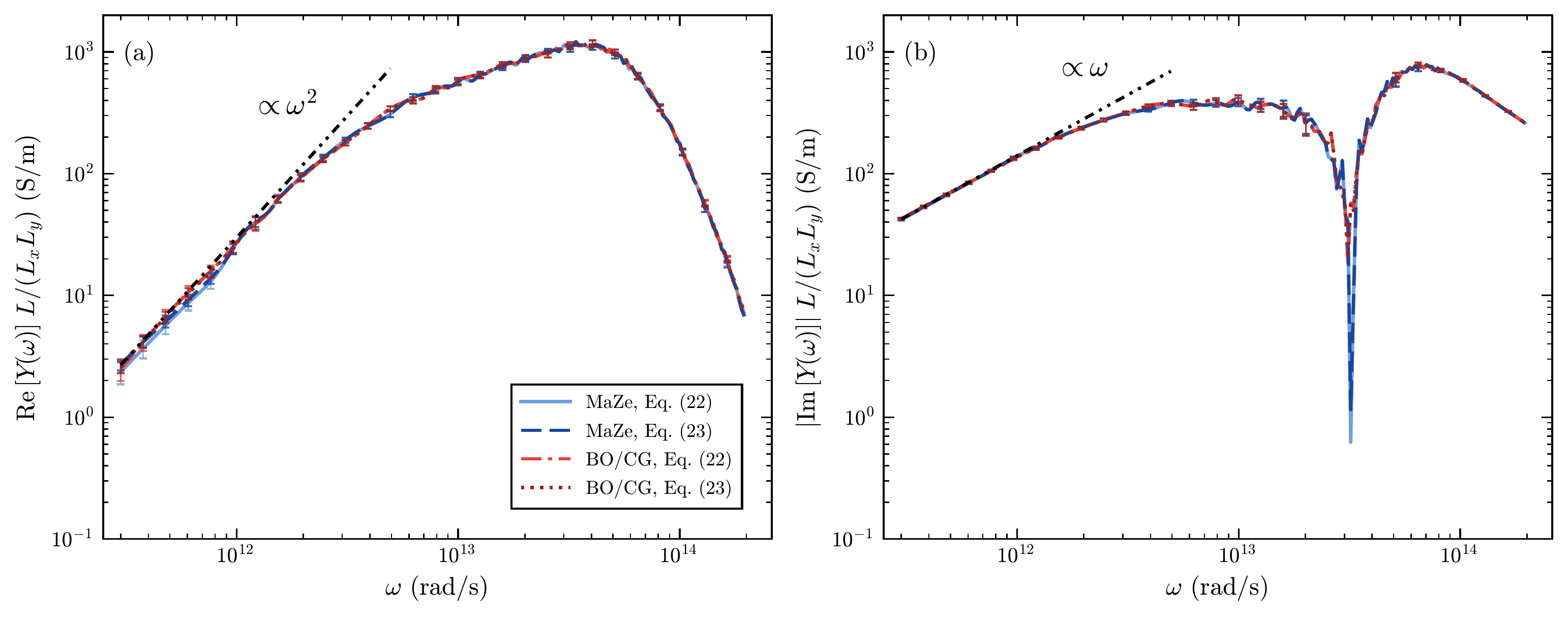}
  \caption{Frequency-dependent admittance of the LiCl--Al cell, (a) real part and (b) absolute value of the imaginary part, normalised by $L_xL_y/L$, computed from the current ACF (Eq.~\eqref{eq:Ycurrent}) and the charge ACF (Eq.~\eqref{eq:Ycharge}) for MaZe and BO/CG. Black dash-dotted-dotted lines: low-frequency power laws $\propto\omega^2$ (a) and $\propto\omega$ (b).  The minimum in (b) near $3\times10^{13}$~rad/s is the sign change of $\mathrm{Im}\,Y$. Error bars are 95\% confidence intervals. They are omitted around the minimum in (b) because $|\mathrm{Im}\,Y|\to0$ at the sign change, where the interval becomes comparable to the value itself and the bars cannot be displayed on a logarithmic axis.}
  \label{fig:admittance}
\end{figure*}

We finally evaluate the admittance using both
Eqs.~\eqref{eq:Ycurrent} and \eqref{eq:Ycharge}, with
one-sided Fourier--Laplace transforms computed
by Filon--Lagrange quadrature with a flat-top Hann taper (details in the SM~\cite{SM}). Fig.~\ref{fig:admittance} compares the real part and the absolute value of the imaginary part from MaZe and BO. The four estimates agree within their $95\%$ confidence intervals across the investigated frequency range. At low frequency, the response
is that of a blocking capacitor in series with a
resistor, with static capacitances
$2.83\pm0.04\,\mu\mathrm{F}/\mathrm{cm}^{2}$ for MaZe
and $2.84\pm0.04\,\mu\mathrm{F}/\mathrm{cm}^{2}$ for BO. The figure also shows that the expected quadratic (for the real part) and linear (for the imaginary part) behavior of this quantity is verified in our calculations. At higher frequency, the response is dominated by the bulk properties of the electrolyte, as signaled by the minimum of $|\mathrm{Im}\,Y|$ at $\omega\approx3\times10^{13}$~rad/s, which corresponds to the change of sign of $\mathrm{Im}\,Y$ and lies close to the maximum of $\mathrm{Re}\,Y$ (a more detailed comparison with the bulk is given in the SM~\cite{SM}). Overall, our results confirm the agreement between the current- and charge-correlation estimators, and between the two dynamics, in both the time and frequency domains.

\vspace{1ex}
{\bf Conclusions} We have shown that the velocities of the constrained auxiliary
variables in MaZe dynamics coincide with the time derivatives of the BO
stationary solutions along the same physical trajectory, and that in Hamiltonian
variables the mass-zero limit is regular, with auxiliary momenta and multipliers
scaling linearly with the inertia and the velocities remaining explicit. The
phase-space compressibility and invariant measure give a statistical
description of these velocities, including the equilibrium
correlation functions and linear-response relations built on them. In
constant-potential molecular dynamics, MaZe and BO give consistent electrode
currents, current autocorrelations and frequency-dependent admittance, with
agreement between estimators based on current and charge correlations.

The correspondence between MaZe and BO is general: it applies whenever auxiliary
variables are defined by smooth stationarity conditions. Examples include induced
dipoles in polarizable force fields, and the related shell and Drude-oscillator models,
whose time derivatives contribute to the polarization current and hence to infrared and
dielectric spectra; fluctuating-charge models, where the charge velocities describe
intramolecular charge-transfer currents; electron densities in orbital-free density
functional theory, whose velocities give access to electronic currents; and
constant-potential electrodes of other geometries, such as nanoporous electrodes in
supercapacitors, where the charge-flow dynamics is the quantity of interest.
For all cases, the results presented in the present work make it possible to interpret and compute the time derivatives of auxiliary BO variables as dynamical variables within a single
constrained-dynamics framework.

\begin{acknowledgments}
This work has received funding from the European Union's Horizon 2020 and Horizon Europe research and innovation programs under the Marie Sk\l{}odowska-Curie (grant agreement no.~101119598, FLUXIONIC), as well as from the European Research Council (grant agreement no.~863473).
\end{acknowledgments}

\bibliography{apssamp}

\clearpage
\onecolumngrid
\setcounter{equation}{0}\setcounter{figure}{0}\setcounter{table}{0}\setcounter{section}{0}
\renewcommand{\theequation}{S\arabic{equation}}
\renewcommand{\thefigure}{S\arabic{figure}}
\renewcommand{\thetable}{S\arabic{table}}
\renewcommand{\thesection}{S\arabic{section}}

\begin{center}
{\large\bfseries Supplemental Material}\\[3pt]
{\bfseries Dynamics and Statistical Mechanics of Born--Oppenheimer Auxiliary Velocities}
\end{center}
\section{\label{app:kappa}Compressibility of the flow}

We start by deriving the phase-space compressibility of the finite-mass constrained
flow in the variables $(r,p,s,\pi)$. The constraint Jacobian and inverse mass matrix are
\begin{equation}
  J=(\Sr,\Ss),\qquad
  W=\begin{pmatrix}\Mmat^{-1}&0\\0&\mu^{-1}\mathbb{1}\end{pmatrix}.
\end{equation}
with $\Ss$ an $(f \times f)$ and $\Sr$ a $(f\times 3N)$ matrix.
Writing $x=(r,s)$ and $v=\dot x=W\Pi$, where $\Pi=(p,\pi)$, the matrix
introduced in Eq.~\eqref{eq:lambda} is
\begin{equation}
  \tilde{\mathbb{Z}}=\mu JWJ^{\mathsf T}
  =\Ss\Ss^{\mathsf T}+\mu\Sr\Mmat^{-1}\Sr^{\mathsf T}.
\end{equation}
\emph{Multipliers.} Differentiating $\sigma(r,s)=0$ twice along the flow gives
$\ddot\sigma_\gamma=v^{\mathsf T}H_\gamma v+(J\ddot x)_\gamma=0$, with
$(H_\gamma)_{ab}=\partial^2\sigma_\gamma/\partial x_a\partial x_b$.
From Eqs.~\eqref{eq:EOMfinite} and $\nabla_sV=\sigma=0$,
$\ddot x=-W(\nabla_rV,0)^{\mathsf T}-WJ^{\mathsf T}\lambda$. Hence
$JWJ^{\mathsf T}\lambda=F$ with
$F_\alpha=v^{\mathsf T}H_\alpha v-(\Sr\Mmat^{-1}\nabla_rV)_\alpha$, as in Eq.~\eqref{eq:Fdef}.
Since $JWJ^{\mathsf T}=\mu^{-1}\tilde{\mathbb{Z}}$,
\begin{equation}
  \lambda=\mu\tilde{\mathbb{Z}}^{-1}F=\mu\gamma ,
\end{equation}

which is Eq.~\eqref{eq:lambda}; $\mu$ multiplies a matrix that is regular at $\mu=0$.
Here $v=W\Pi=(\Mmat^{-1}p,\pi/\mu)$ is regarded as a function of the independent
variables $(p,\pi)$; on the constraint surface $\pi/\mu=\hat\pi$,
Eq.~\eqref{eq:scaledmomentum}.

The phase-space divergence receives contributions only from the momentum
equations:
\begin{equation}
  \kappa_\mu
  =-\mu\sum_{\beta}\left[
    \sum_i(\Sr)_{\beta i}\frac{\partial\gamma_\beta}{\partial p_i}
    +\sum_\alpha(\Ss)_{\beta\alpha}
      \frac{\partial\gamma_\beta}{\partial\pi_\alpha}
  \right].
\end{equation}
Define
\begin{equation}
  D_\beta=\mu\sum_i(\Sr)_{\beta i}\frac{\partial}{\partial p_i}
  +\mu\sum_\alpha(\Ss)_{\beta\alpha}
    \frac{\partial}{\partial\pi_\alpha}.
\end{equation}
Because $\tilde{\mathbb{Z}}$ is independent of the momenta,
\begin{equation}
  \kappa_\mu=-\sum_{\beta\gamma}
    (\tilde{\mathbb{Z}}^{-1})_{\beta\gamma}D_\beta F_\gamma.
  \label{eq:kappa-DF}
\end{equation}

Let $(H_\gamma)_{ab}=\partial^2\sigma_\gamma/
(\partial x_a\partial x_b)$. From Eq.~\eqref{eq:Fdef},
\begin{equation}
  F_\gamma=v^{\mathsf T}H_\gamma v
  -(\Sr\Mmat^{-1}\nabla_r V)_\gamma.
\end{equation}
The second term is independent of the momenta, while
\begin{equation}
  D_\beta v_a=\mu(WJ^{\mathsf T})_{a\beta},\qquad
  \dot J_{\gamma a}=\sum_b(H_\gamma)_{ab}v_b.
\end{equation}
It follows that
\begin{equation}
  D_\beta F_\gamma
  =2\mu\sum_{ab}(H_\gamma)_{ab}v_b
    (WJ^{\mathsf T})_{a\beta}
  =2\mu(\dot JWJ^{\mathsf T})_{\gamma\beta}.
\end{equation}
Thus,
\begin{equation}
  \kappa_\mu=-2\mu\Tr\bigl(
    \tilde{\mathbb{Z}}^{-1}\dot JWJ^{\mathsf T}\bigr).
\end{equation}
Differentiating $\tilde{\mathbb{Z}}=\mu JWJ^{\mathsf T}$ at fixed $\mu$ gives
\begin{equation}
  \dot{\tilde{\mathbb{Z}}}=\mu\bigl(
    \dot JWJ^{\mathsf T}+JW\dot J^{\mathsf T}\bigr).
\end{equation}
Using the symmetry of $W$ and $\tilde{\mathbb{Z}}^{-1}$ and the cyclic
property of the trace, we obtain
\begin{equation}
  \kappa_\mu
    =-\Tr\bigl(\tilde{\mathbb{Z}}^{-1}\dot{\tilde{\mathbb{Z}}}\bigr)
    =-\frac{\rd}{\rd t}\ln|\tilde{\mathbb{Z}}|
  =-\frac{\rd}{\rd t}\ln\sqrt g,
  \label{eq:compressibilityS}
\end{equation}
where the last equality uses precisely the metric factor of
Eq.~\eqref{eq:metric},
\begin{equation}
  \sqrt g=|\tilde{\mathbb{Z}}|.
\end{equation}
The momentum derivatives above are taken at fixed coordinates,
treating $p$ and $\pi$ as independent variables; the velocity
constraint is imposed through $\delta(\dot\sigma)$ in the
constrained phase-space measure.

If $\Ss$ is nonsingular and the constrained velocities remain finite,
the mass-zero limit is
\begin{equation}
  \kappa_0=-\frac{\rd}{\rd t}
    \ln|\Ss\Ss^{\mathsf T}|.
  \label{eq:kappa-zero}
\end{equation}
For the constant-potential model, $\Ss$ is independent of time and hence $\kappa_0=0$.

\section{Regular mass-zero limit of the invariant measure}

The reduction of the constrained equilibrium measure to the Born--Oppenheimer (BO)
density in the mass-zero limit was established in previous
work~\cite{Bonella2020,Coretti2020}. We include it in this work for completeness, providing a different derivation
recast in the Hamiltonian variables used here, where it follows from the regular scaled momenta.
We show how the average of Eq.~\eqref{eq:average} reduces to the BO density as $\mu\to0$. Change variables to $\pi=\mu\hat\pi$, so that
$\rd\pi=\mu^{f}\rd\hat\pi$ (the factor $\mu^f$ is common to numerator and denominator
and cancels). On the constraint surface the velocity constraint is
\begin{equation}
  \delta(\dot\sigma)=\delta(\Sr\Mmat^{-1}p+\Ss\hat\pi)
  =|\Ss|^{-1}\,\delta\big(\hat\pi-\hat\pi_*(r,p)\big),
  \qquad
  \hat\pi_*\equiv-\Ss^{-1}\Sr\Mmat^{-1}p ,
\end{equation}
with no dependence on $\mu$. Integrating over $\hat\pi$ and then over $s$ with
$\int\rd s\,\delta(\sigma(r,s))\,g(s)=g(s^*)/|\Ss(r,s^*)|$ gives, at finite $\mu$,
\begin{equation}
  \langle A\rangle_\mu=\frac1{\mathcal N_\mu}\int\rd r\,\rd p\;
  \frac{|\tilde{\mathbb{Z}}|}{|\Ss|^{2}}\;P_\mu\,A\Big|_{s=s^*(r),\ \hat\pi=\hat\pi_*(r,p)} .
\end{equation}
By the matrix determinant lemma,
$|\tilde{\mathbb{Z}}|=|\Ss\Ss^{\mathsf T}|\,|\mathbb 1+\mu T|$ with
$T=(\Ss\Ss^{\mathsf T})^{-1}\Sr\Mmat^{-1}\Sr^{\mathsf T}$, so that
$|\tilde{\mathbb{Z}}|/|\Ss|^2=|\mathbb 1+\mu T|\to1$ as $\mu\to0$. In the limit the physical
variables $(r,p)$ are therefore distributed according to $P_0(r,p,s^*(r),\hat\pi_*)$,
the BO density for the chosen ensemble, with no residual metric factor, and the
auxiliary velocity is the slaved value $\hat\pi_*=\dot s^*$. The same steps applied
to Eq.~\eqref{eq:averagezero}, where $|\tilde{\mathbb{Z}}|\to|\Ss\Ss^{\mathsf T}|$, give the weight
$|\Ss|\,\delta(\sigma)\,\delta(\hat\pi-\hat\pi_*)$ quoted in the main text. For finite $\mu$
the metric contributes $|\mathbb 1+\mu T|$, where the $O(\mu)$ term does not vanish identically for generic $\Sr$. Furthermore, the Hamiltonian contains a modified inertia for the physical variables, reading $H_\mu=\frac12p^{\mathsf T}[\mathbb{1}-\mu\Sigma_s^{-1}\Sigma_r]\Mmat^{-1}p+V(r,s)$.

\section{Details of the constant-potential system}
The constraints are $\sigma_\alpha=\partial U/\partial Q_\alpha$ ($\alpha=1,\dots,M$) and $\sigma_{M+1}=\partial U/\partial\nu$, with
\begin{equation}
U(r,Q,\nu)=U_C-\sum_{\alpha=1}^{M}\psi_\alpha Q_\alpha+\nu\sum_{\alpha=1}^{M}Q_\alpha ,
\end{equation}
and $s=(Q_1,\dots,Q_M,\nu)$. The matrices $\Sigma_s=\partial\sigma/\partial s$ and $\Sigma_r=\partial\sigma/\partial r$ only require derivatives of the Coulomb energy $U_C$. With an Ewald-type splitting, $U_C$ separates into ion--ion, ion--electrode and electrode--electrode contributions,
$U_C = U_C^{qq}(r) + U_C^{qQ}(r,Q) + U_C^{QQ}(Q)$. In their bare (non-periodic) form these read
\begin{align}
  U_C^{QQ}(Q) &= \sum_{\alpha}\sum_{\beta<\alpha}
              \frac{Q_\alpha Q_\beta}{R_{\alpha\beta}}
              \operatorname{erf}\!\big[\eta_{\alpha\beta}\,R_{\alpha\beta}\big]
            + \sum_{\alpha}\frac{\eta_\alpha}{\sqrt{2\pi}}\,Q_\alpha^2 ,
              \nonumber\\
  U_C^{qQ}(r,Q) &= \sum_{\alpha}\sum_i
              \frac{Q_\alpha q_i}{|R_\alpha-r_i|}
              \operatorname{erf}\!\big[\eta_\alpha\,|R_\alpha-r_i|\big] ,
              \nonumber\\
  U_C^{qq}(r) &= \sum_{i}\sum_{j<i}\frac{q_i q_j}{r_{ij}} ,
  \label{eq:UC}
\end{align}
with $q_i$ the ionic charges, $R_\alpha$ the (fixed) electrode-site positions,
$R_{\alpha\beta}=|R_\alpha-R_\beta|$, $r_{ij}=|r_i-r_j|$, $\eta_\alpha$ the inverse width of the Gaussian charge on electrode site $\alpha$ (charge density $\propto e^{-\eta_\alpha^2r^2}$), and
$\eta_{\alpha\beta}=\eta_\alpha\eta_\beta/\sqrt{\eta_\alpha^2+\eta_\beta^2}$ ($=\eta/\sqrt2$ for equal widths $\eta$).
Here and in the following, Coulomb interactions are written in atomic units ($4\pi\varepsilon_0=1$).
The periodic (Ewald) form of these terms is given below, with $\kappa_c$ the Ewald splitting parameter.
Since $U_C^{qq}$ does not depend on $Q$, $U_C^{qQ}$ is linear in $Q$, and $U_C^{QQ}$ does not depend on $r$, $\Sigma_s$ only involves $U_C^{QQ}$, and $\Sigma_r$ only involves $U_C^{qQ}$.

\subsection*{$\Sigma_s$: derivatives with respect to $s_i$}
The auxiliary Hessian is the bordered electrode--electrode Coulomb matrix
\begin{equation}
  \Ss =
  \begin{pmatrix}
    U_{11} & \cdots & U_{1M} & 1\\[2pt]
    \vdots & \ddots & \vdots & \vdots\\[2pt]
    U_{M1} & \cdots & U_{MM} & 1\\[2pt]
    1 & \cdots & 1 & 0
  \end{pmatrix},
  \qquad
  U_{\alpha\beta}\equiv\frac{\partial^2 U_C}{\partial Q_\alpha\partial Q_\beta},
  \label{eq:SigmaS}
\end{equation}
whose nontrivial block $U_{\alpha\beta}$ depends only
on the fixed electrode geometry $\{R_\alpha\}$, on the box and on $\eta_\alpha$, $\kappa_c$.

\emph{Electrode--electrode contribution $U_C^{QQ}$.}
It is split into short-range, long-range ($k=0$ and $k\neq0$), and self-energy terms,
\begin{align}
U_C^{QQ} &= U_{sr}^{QQ} + U_{lr,0}^{QQ} + U_{lr,*}^{QQ} + U_{self}^{QQ}.
\end{align}
The double sums run over ordered pairs $(\alpha,\beta)$, which is why the short-range term carries a factor $\tfrac12$ and the Hessians below differ from the prefactors of $U$ accordingly.

\emph{Short range.} With $\rho_{\alpha\beta}\equiv|\vec R_{\alpha\beta}+\vec m|$ ($\vec m$ the in-plane lattice vector labelled by $n$; the prime excludes $n=0$ when $\alpha=\beta$) and $\Delta_{\alpha\beta}(\rho)\equiv\operatorname{erfc}[\kappa_c\rho]-\operatorname{erfc}[\eta_{\alpha\beta}\rho]$,
\begin{align}
U_{sr}^{QQ} &= \frac12\sum_{\alpha,\beta}^{M}{\sum_n}'\,
\frac{Q_\alpha Q_\beta}{\rho_{\alpha\beta}}\,\Delta_{\alpha\beta}(\rho_{\alpha\beta}),\\
\frac{\partial^2 U_{sr}^{QQ}}{\partial Q_\alpha\partial Q_\beta} &=
{\sum_n}'\,\frac{\Delta_{\alpha\beta}(\rho_{\alpha\beta})}{\rho_{\alpha\beta}}.
\end{align}

\emph{Long range, $k=0$.} With
$\mathcal G(\zeta)\equiv e^{-\kappa_c^2\zeta^2}/\kappa_c+\sqrt\pi\,\zeta\operatorname{erf}[\kappa_c\zeta]$,
\begin{align}
U_{lr,0}^{QQ} &= -\frac{\sqrt\pi}{L_xL_y}\sum_{\alpha,\beta}^{M}Q_\alpha Q_\beta\,\mathcal G(Z_{\alpha\beta}),\\
\frac{\partial^2 U_{lr,0}^{QQ}}{\partial Q_\alpha\partial Q_\beta} &=
-\frac{2\sqrt\pi}{L_xL_y}\,\mathcal G(Z_{\alpha\beta}).
\end{align}

\emph{Long range, $k\neq0$.} With
$\mathcal K(h,u)\equiv e^{-(h^2+u^2)/4\kappa_c^2}/(h^2+u^2)$, which is even in $(\tilde h,u)$ so that the sums below are real,
\begin{align}
U_{lr,*}^{QQ} &= \frac{1}{L_xL_y}\sum_{\alpha,\beta}^{M}Q_\alpha Q_\beta
\int_{-\infty}^{\infty}\!du\sum_{h\neq0}\mathcal K(h,u)
\times\exp\!\big[i(\tilde h\cdot\tilde R_{\alpha\beta}+uZ_{\alpha\beta})\big],\\
\frac{\partial^2 U_{lr,*}^{QQ}}{\partial Q_\alpha\partial Q_\beta} &=
\frac{2}{L_xL_y}\int_{-\infty}^{\infty}\!du\sum_{h\neq0}\mathcal K(h,u)
\times\cos\!\big(\tilde h\cdot\tilde R_{\alpha\beta}+uZ_{\alpha\beta}\big).
\end{align}

\emph{Self energy.}
\begin{align}
U_{self}^{QQ} &= \sum_{\alpha=1}^{M}\left(\frac{\eta_\alpha}{\sqrt{2\pi}} - \frac{\kappa_c}{\sqrt\pi}\right)Q_\alpha^2,\\
\frac{\partial^2 U_{self}^{QQ}}{\partial Q_\alpha^2} &= 2\left(\frac{\eta_\alpha}{\sqrt{2\pi}} - \frac{\kappa_c}{\sqrt\pi}\right).
\end{align}

\subsection*{$\Sigma_r$: derivatives with respect to $r_i$}
The elements of $\Sigma_r$ are
\begin{align}
\frac{\partial\sigma_\alpha}{\partial r_i} &= \frac{\partial}{\partial r_i}\frac{\partial U}{\partial Q_\alpha}
 = \frac{\partial^2 U_C}{\partial r_i\,\partial Q_\alpha},\\
\frac{\partial\sigma_{M+1}}{\partial r_i} &= \frac{\partial}{\partial r_i}\frac{\partial U}{\partial\nu} = 0,
\end{align}
because $\sigma_{M+1}=\sum_\alpha Q_\alpha$ does not depend on $r$.

\emph{Electrode--ion contribution $U_C^{qQ}$.}
\begin{equation}
U_C^{qQ} = U_{sr}^{qQ} + U_{lr,0}^{qQ} + U_{lr,*}^{qQ}.
\end{equation}
Each ion--electrode pair appears twice in the sum over ordered pairs of all particles, which gives the factors below.

\emph{Short range.} With $\vec d\equiv\vec r_i-\vec R_\alpha+\vec m$, $d\equiv|\vec d|$ and $\Delta_\alpha(d)\equiv\operatorname{erfc}[\kappa_c d]-\operatorname{erfc}[\eta_\alpha d]$,
\begin{align}
U_{sr}^{qQ} &= \sum_i^{N}\sum_\alpha^{M}\sum_n\,\frac{q_iQ_\alpha}{d}\,\Delta_\alpha(d),\\
\frac{\partial^2 U_{sr}^{qQ}}{\partial \vec r_i\,\partial Q_\alpha} &= \sum_n\,\frac{q_i\vec d}{d^2}
\Big[-\frac{\Delta_\alpha(d)}{d}-\frac{2\kappa_c}{\sqrt\pi}e^{-\kappa_c^2d^2}
+\frac{2\eta_\alpha}{\sqrt\pi}e^{-\eta_\alpha^2d^2}\Big].
\end{align}

\emph{Long range, $k=0$.} Using $\mathcal G$ as above,
\begin{equation}
U_{lr,0}^{qQ} = -\frac{2\sqrt\pi}{L_xL_y}\sum_i^{N}\sum_\alpha^{M}q_iQ_\alpha\,\mathcal G(z_i-Z_\alpha),
\end{equation}
which does not depend on $x_i,y_i$, and, since $\mathcal G'(\zeta)=\sqrt\pi\operatorname{erf}[\kappa_c\zeta]$,
\begin{equation}
\frac{\partial^2 U_{lr,0}^{qQ}}{\partial z_i\,\partial Q_\alpha} =
-\frac{2\pi}{L_xL_y}\,q_i\operatorname{erf}\big[\kappa_c(z_i-Z_\alpha)\big].
\end{equation}

\emph{Long range, $k\neq0$.} With $\tilde R_\alpha=(X_\alpha,Y_\alpha)$, $\tilde r_i=(x_i,y_i)$ and $\Phi_{i\alpha}\equiv\tilde h\cdot(\tilde R_\alpha-\tilde r_i)+u(Z_\alpha-z_i)$,
\begin{align}
U_{lr,*}^{qQ} &= \frac{2}{L_xL_y}\sum_i^{N}\sum_\alpha^{M}q_iQ_\alpha
\int_{-\infty}^{\infty}\!du\sum_{h\neq0}\mathcal K(h,u)\,\cos\Phi_{i\alpha},\\
\frac{\partial^2 U_{lr,*}^{qQ}}{\partial r_i\,\partial Q_\alpha} &=
\frac{2q_i}{L_xL_y}\int_{-\infty}^{\infty}\!du\sum_{h\neq0}\mathcal K(h,u)\,\sin\Phi_{i\alpha}
\times
\begin{cases}
\tilde h & \text{for } x,y,\\
u & \text{for } z.
\end{cases}
\end{align}

\section{\label{app:protocol}Simulation protocol and comparison with the bulk}

Following Ref.~\citenum{Coretti2020}, all simulations in this work, including the bulk reference described below, use the same protocol for equilibration and production. After an NVT equilibration at $1500\,\mathrm{K}$ with a Nos\'e--Hoover chain thermostat, we perform NVE production runs of $1\,\mathrm{ns}$ with a $1\,\mathrm{fs}$ timestep. Independent MaZe and Born--Oppenheimer runs of the cell are started from the same equilibrated configuration, with ion velocities redrawn from a Maxwell--Boltzmann distribution at $1500\,\mathrm{K}$, and are used to collect statistics. The one-sided Fourier--Laplace transforms in Eqs.~\eqref{eq:Ycurrent} and \eqref{eq:Ycharge} were evaluated by Filon--Lagrange quadrature~\cite{Pireddu2023}, with a flat-top Hann taper applied to suppress long-time statistical noise~\cite{Pireddu2024}. We use two windows: $5$~ps (flat for the first $1.5$~ps) above $1.5\times10^{12}$~rad/s, and $20$~ps (flat for the first $5$~ps) below. Low frequencies are determined by the long-time part of the autocorrelation functions, where the signal has decayed and the statistical noise dominates, the larger window corrects for this. The expected low-frequency behaviour follows from the blocking capacitor in series with a resistor~\cite{Pireddu2023}, $Y(\omega)=i\omega C/(1+i\omega RC)$, which gives $\mathrm{Im}\,Y\simeq\omega C$ and $\mathrm{Re}\,Y\simeq RC^{2}\omega^{2}$ for $\omega RC\ll1$. In the current autocorrelation, $\mathrm{Re}\,Y\propto\omega^2$ is the small remainder of a cancellation between positive and negative parts (the blocking electrodes imply a vanishing zero-frequency conductance, $\int_0^\infty\langle\delta I(0)\delta I(t)\rangle\,dt=0$), which a short taper breaks, so a longer window is needed; at higher frequency the longer window only adds noise. Each 1 ns trajectory was divided into ten non-overlapping blocks of 100 ps. For each method (MaZe and BO/CG) we pooled the blocks of seven independent trajectories, giving 70 blocks per method.

To interpret the response of the cell over the whole frequency range, we compared it with a simulation of the molten salt in the absence of electrodes. The bulk reference is a fully periodic box without electrodes, containing the same $500$ Li$^+$ and $500$ Cl$^-$ ions and the same interaction potential as the cell, in a box of $24.30\times24.30\times46.87\,\text{\AA}^3$.

To compare the cell with the bulk we need to define an effective slab thickness for the liquid. For dielectric liquids this can be done by locating a dielectric dividing surface (DDS) at each interface, which separates the electrode from a liquid with a well-defined bulk dielectric response~\cite{Pireddu2023}. In a dilute electrolyte, the DDS of the pure solvent has been used to define the effective slab thickness, even though in that case the response arises not only from the polar solvent but also from ionic transport~\cite{Pireddu2024}. For a conducting liquid such as molten LiCl this construction cannot be used: the liquid screens the electrodes, so the polarization is not localised in the bulk of the slab and no dielectric dividing surface can be defined. We therefore introduce an effective thickness $d_\mathrm{eff}=L-2w$,  where $L$ is the distance between the innermost electrode planes and $w$ is the width of the ion-depleted layer at each interface, taken from the Gibbs dividing surface of the Li$^+$ and Cl$^-$ density profiles. In our simulations $w=1.34$~\AA, so that $d_\mathrm{eff}=47.15$~\AA\ for $L=49.83$~\AA. Varying $w$ between $1$ and $2$~\AA\ changes our results by less than $3\%$, and neither the bulk resistivity nor the static capacitance depends on this choice. The frequency-dependent conductivity of the bulk is obtained from the Green--Kubo relation
\begin{equation}
 \sigma(\omega)=\frac{\beta}{3\mathcal{V}}\sum_{\alpha=x,y,z}\int_0^\infty \langle J_\alpha(t)J_\alpha(0)\rangle e^{-i\omega t}\,dt,
 \qquad \mathbf{J}(t)=\sum_i q_i \mathbf{v}_i(t),
\end{equation}
where $\mathcal{V}$ is the volume of the bulk simulation box, $q_i$ and $\mathbf{v}_i$ are the charge and velocity of ion $i$, and $\beta=1/k_BT$, with the same procedure used for the cell.

\begin{figure*}[t]
  \centering
    \includegraphics[width=\linewidth]{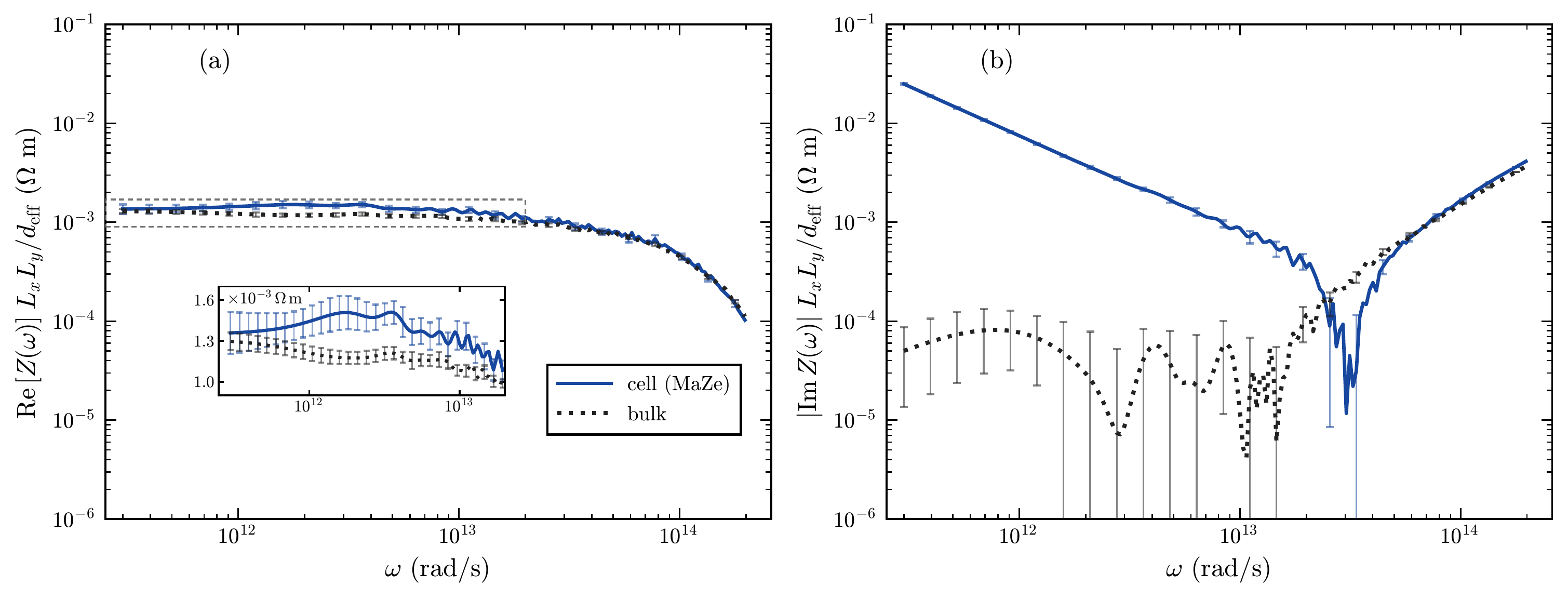}
  \caption{Impedance of the LiCl--Al cell (MaZe) compared with bulk LiCl, in a slab of effective thickness $d_{\rm eff}$ ($d_{\rm eff}=L-2w$, see the text): (a) $\mathrm{Re}\,Z$ and (b) $|\mathrm{Im}\,Z|$, both multiplied by $L_xL_y/d_{\rm eff}$ so that they can be compared directly with the bulk resistivity. The inset in (a) magnifies the low-frequency region on a linear scale. At low frequency $|\mathrm{Im}\,Z|$ of the cell diverges as $1/\omega$ because of the interfacial capacitance, which is absent in the bulk; above $\omega_{RC}$ the two curves coincide. The minima of $|\mathrm{Im}\,Z|$ near $3\times10^{13}$~rad/s correspond to the sign change of $\mathrm{Im}\,Z$. Error bars are 95\% confidence intervals. The bulk $\mathrm{Im}\,Z$ is small and compatible with zero within its $95\%$ confidence interval over most of the range between $10^{12}$ and $1.6\times10^{13}$~rad/s, which is why its error bars in (b) extend down past the lower edge of the logarithmic axis.}
  \label{fig:bulk_Z}
\end{figure*}

Fig.~\ref{fig:bulk_Z} compares the cell impedance rescaled by the geometry of the liquid slab, $Z(\omega)\,A_\mathrm{el}/d_\mathrm{eff}$, with the bulk resistivity $1/\sigma(\omega)$ of molten LiCl, where $A_\mathrm{el}=L_xL_y$ is the electrode area, $d_\mathrm{eff}$ is the effective thickness of the liquid slab defined above, and $\sigma(\omega)$ is the frequency-dependent conductivity of the bulk liquid obtained from the separate bulk simulation. The comparison follows from the picture of the cell as the bulk liquid in series with the interfacial capacitance~\cite{Pireddu2024},
\begin{equation}
  Z_\mathrm{cell}(\omega)\simeq \frac{d_\mathrm{eff}}{A_\mathrm{el}\,\sigma(\omega)}+\frac{1}{i\omega C_\mathrm{int}} ,
\end{equation}
so that $Z_\mathrm{cell}A_\mathrm{el}/d_\mathrm{eff}\simeq 1/\sigma(\omega)$ wherever the interfacial term $1/(\omega C_\mathrm{int})$ is negligible compared with the bulk term. The capacitance $C_\mathrm{int}$ arises only from the electrode--electrolyte interfaces and is absent in the bulk liquid, which has no electrodes and whose response is entirely contained in $\sigma(\omega)$.

The real part is set by the bulk: $\mathrm{Re}\,Z_{\rm cell}$ equals $1.4\times10^{-3}\,\Omega\,\mathrm{m}$ at low frequency, within about $10\%$ of the bulk value $1.3\times10^{-3}\,\Omega\,\mathrm{m}$, with an interfacial excess $\mathrm{Re}\,Z_{\rm int}=\mathrm{Re}(Z_{\rm cell}-Z_{\rm bulk})$ of about $15\%$ below $8\times10^{12}$~rad/s. This indicates that the resistance of the cell is dominated by the bulk electrolyte. The imaginary part is instead dominated by the interface at low frequency, where $|\mathrm{Im}\,Z_{\rm cell}|\simeq 1/(\omega C_{\rm int})$ reaches $2.5\times10^{-2}\,\Omega\,\mathrm{m}$ at $\omega=3\times10^{11}$~rad/s, more than two orders of magnitude above the bulk value at the same frequency, which is of order $10^{-4}\,\Omega\,\mathrm{m}$ or below. The bulk $\mathrm{Im}\,Z$ is statistically indistinguishable from zero below $\approx 1.6\times10^{13}$~rad/s ($95\%$ CI), as expected since the bulk behaves as a plain resistor at low frequency, so its impedance is real. The minimum of $|\mathrm{Im}\,Z_{\rm cell}|$ at $\omega\approx3\times10^{13}$~rad/s corresponds to the sign change of $\mathrm{Im}\,Z_{\rm cell}$ (and of $\mathrm{Im}\,Y$). The cell and bulk impedances merge for $\omega\gtrsim3\times10^{13}$~rad/s, i.e.\ above the sign change of $\mathrm{Im}\,Y(\omega)$ (and of $-\mathrm{Im}\,Z_{\rm cell}$), when the capacitive term is smaller than the bulk contribution.

\end{document}